\documentclass[aps,preprint,prd,showpacs,nofootinbib]{revtex4}

\usepackage{latexsym}
\usepackage{CJK}
\usepackage{amsmath}
\usepackage{graphicx}
\usepackage{dcolumn}
\usepackage{bm}
\usepackage{amssymb}
\usepackage{latexsym}
\usepackage{color}
\usepackage[colorlinks,linkcolor=red,anchorcolor=green,citecolor=blue]{hyperref}

\def\be{\begin{equation}}
\def\ee{\end{equation}}
\def\ba{\begin{eqnarray}}
\def\ea{\end{eqnarray}}

\def\nn{\nonumber}

\begin{document}

\title{Oscillating modulation to B-mode polarization from varying propagating speed of primordial gravitational waves}

\author{Yong Cai$^{1,}$\footnote{Email: caiyong13@mails.ucas.ac.cn}}
\author{Yu-Tong Wang$^{1}$\footnote{Email: wangyutong12@mails.ucas.ac.cn}}
\author{Yun-Song Piao$^{1,2}$\footnote{Email: yspiao@ucas.ac.cn}}

\affiliation{$^1$ School of Physics, University of Chinese Academy of
Sciences, Beijing 100049, China}

\affiliation{$^2$ State Key Laboratory of Theoretical Physics, Institute of Theoretical Physics, \\
Chinese Academy of Sciences, P.O. Box 2735, Beijing 100190, China}

\begin{abstract}

In low-energy effective string theory and modified gravity
theories, the propagating speed $c_T$ of primordial gravitational
waves may deviate from unity. We find that the step-like variation
of $c_T$ during slow-roll inflation may result in an oscillating
modulation to the B-mode polarization spectrum, which can hardly
be imitated by adjusting other cosmological
parameters, and the intensity of the modulation is determined by
the dynamics of $c_T$. Thus provided that the foreground
contribution is under control, high-precision CMB polarization
observations will be able to put tight constraint on the variation
of $c_T$, and so the corresponding theories.

\end{abstract}
\maketitle

\section{Introduction}

Inflation, as the paradigm of the early universe, has not only
solved a lot of fine-tuning problems of the big bang theory, but
also predicted the primordial scalar and tensor perturbations. The
primordial tensor perturbations, i.e. primordial gravitational
waves (GWs) \cite{Gri:1974}\cite{Sta:1979}\cite{Rub:1982}, have
arisen great attentions after the BICEP2 collaboration's
announcement of the detection of B-mode signal in the CMB (around
$l\sim 80$) \cite{Ade:2014xna}, which was interpreted by them as
the imprint of the primordial GWs, though this result is doubtful
due to the foregrounds of polarized dust emissions
\cite{Adam:2014bub}\cite{Ade:2015tva}, see also \cite{Mortonson:2014bja}.

The detection of primordial GWs would verify general relativity
(GR) and strengthen our confidence in inflation and quantum gravity\cite{Ashoorioon:2012kh}, and also put more
constrains on inflation models and modified gravity at the same
time.
Besides the CMB experiments which mainly aimed at detecting low
frequencies ($10^{-17}- 10^{-15}$ Hz) GWs,  many experiments
relate to higher frequencies based on other methods, such as
pulsar timing array ($10^{-9}- 10^{-8}$ Hz), laser
interferometer detectors ($10^{-4}- 10^4$ Hz), will be carried out
in the coming decades. However, since the amplitude of the GWs
would stay constant after they are stretched outside the horizon,
and decrease with the expansion of the universe after they reenter
the horizon, the primordial GWs with longer wavelength provide the
most of opportunities for the detection \cite{Grishchuk:2007uz}.
Therefore, the CMB observations, especially the CMB B-mode
detections, are still the most promising experiments to detect the
primordial GWs if they actually exist.

Einstein's GR is the most accepted theory of gravity. However, it
might be required to modify when dealing with the inflation in the
primordial universe and the accelerated expansion of the current
universe. During the matter and radiation dominated era, modified
gravity has several effects on the CMB spectra, such as the
lensing contribution to B-modes
\cite{Lewis:2006fu} and the variation of propagating speed $c_T$
of primordial GWs \cite{Amendola:2014wma}\cite{Raveri:2014eea}, we
are especially interested in the latter in this paper, see
e.g.\cite{Piao:2006ja} for the case with the scalar perturbation.
In GR, graviton is massless and propagates along the null
geodesics, so the propagating speed $c_T$ of GWs is naturally set
to be unity, i.e. the speed of light. But in modified gravity,
e.g., the low-energy effective string theory with high-order
corrections
\cite{Met:1987}\cite{Antoniadis:1993jc}\cite{Cartier:1999vk}\cite{Cartier:2001is}\cite{Piao:2003hh},
and also modified Gauss-Bonnet gravity \cite{Nojiri:2005jg}, and
generalized Galileon (Horndeski theory \cite{Horndeski:1974wa})
\cite{Amendola:1993uh}\cite{Deffayet:2011gz}\cite{Kobayashi:2011nu}\cite{Liu:2011ns},
and beyond Horndeski theories
\cite{Gleyzes:2014dya},\cite{Gao:2014soa}, and the
effective theory of fluids at next-to-leading order in derivatives
(e.g.\cite{Ballesteros:2014sxa}), $c_T$ might deviate from unity.
Because the value of $c_T$ determines the time of horizon crossing
of GWs, during the recombination the change of $c_T$ can result in
a shift of the peak position of the primordial B-modes, see
\cite{Amendola:2014wma}\cite{Raveri:2014eea}, for example.

In this paper, we focus on the effect of the variation of $c_T$
during slow-roll inflation on the CMB B-mode polarization, and
show how it offers a distinct way to test the modified gravity
theories. We find that the step-like variation of $c_T$ may result
in an oscillating modulation to the B-mode polarization spectrum,
which can hardly be imitated by adjusting other cosmological
parameters. The intensity of the modulation is determined by the
ratio of $c_T$ before and after the variation, which depends on
the dynamics of theoretical models.  This oscillating
modulation is so rich in feature that it may easily be
discriminated from the variation of other parameters or other
features. Thus provided that the foreground contribution is under
control, high-precision CMB polarization observations will be able
to put tight constraint on the variation of $c_T$, and so the
corresponding theories.

\section{Oscillating spectrum of primordial GWs}


We begin with the action for the GWs mode $h_{ij}$,
e.g.\cite{Cartier:2001is}\cite{Kobayashi:2011nu} \ba S_{(2)}=\int
d\tau d^3x\, {a^2 Q_T\over 8} \left[ h^{\prime
2}_{ij}-c_T^2(\vec{\nabla} h_{ij})^2 \right],\label{action2} \ea
where $h_{ij}$ obeys $\partial _i h_{ij}=0$ and $h_{ii}=0$, $Q_T$
is regarded as effective Planck scale $M_{P,eff}^2(\tau)$, $c_T$
is the propagating speed of primordial GWs, and the prime is the
derivative with respect to the conformal time $\tau$,
$d\tau=dt/a$. During slow-roll inflation, the slow-roll parameter
$ \epsilon=-{\dot{H}}/{H^2}\ll 1$, as well as
\be \epsilon_{Q}={{\dot Q}_T\over H Q_T}\ll 1,\,\,\,\,\, s={{\dot
c}_T\over Hc_T}\ll 1\ee are required, e.g., see
\cite{DeFelice:2014bma} for a recent study.

Here, we will mainly focus on the effect of varying $c_T$, i.e.,
the condition $s\ll 1$ might be broke at some point, on primordial
GWs spectrum. Noting that the effects of varying sound speed
of primordial scalar perturbations on the scalar power spectrum
have been investigated in
\cite{Nakashima:2010sa}\cite{Firouzjahi:2014fda}. We do not get
entangled with the details of (\ref{action2}) and the evolution of
background, and will assume that the background is the slow-roll
inflation, which is not affected by the variation of $c_T$, and
set $Q_T$ constant and $M_P^2=1$. We will discuss a possibility of
such a case in Appendix A. In addition, there may be a mass term
\cite{Dubovsky:2010pe}\cite{Gumrukcuoglu:2012wt} in
(\ref{action2}), which might also be time-dependent
\cite{Huang:2012pe}, but we will not involve it.

We can expand $h_{ij}$ into Fourier series as $h_{ij}(\tau,\mathbf{x})=\int \frac{d^3k}{(2\pi)^{3} }e^{-i\mathbf{k}\cdot \mathbf{x}}\hat{h}_{ij}(\tau,\mathbf{k})$, where 
\be \hat{h}_{ij}(\tau,\mathbf{k})=\sum_{\lambda=+,\times}\left[ h_{\lambda}(\tau,k)a_{\lambda}(\mathbf{k})
+h_{\lambda}^*(\tau,-k)a_{\lambda}^{\dag}(-\mathbf{k}) \right]
\epsilon^{(\lambda)}_{ij}(\mathbf{k}), \label{hij}\ee
where the polarization tensors
$\epsilon_{ij}^{(\lambda)}(\mathbf{k})$ are defined by
$k_{j}\epsilon_{ij}^{(\lambda)}(\mathbf{k})=0$,
$\epsilon_{ii}^{(\lambda)}(\mathbf{k})=0$, which satisfy
$\epsilon_{ij}^{(\lambda)}(\mathbf{k})
\epsilon_{ij}^{*(\lambda^{\prime}) }(\mathbf{k})=\delta_{\lambda
\lambda^{\prime} }$, $\epsilon_{ij}^{*(\lambda)
}(\mathbf{k})=\epsilon_{ij}^{(\lambda) }(-\mathbf{k})$, the
commutation relation for the annihilation and creation operators
$a_{\lambda}(\mathbf{k})$ and
$a^{\dag}_{\lambda}(\mathbf{k}^{\prime})$ is $[
a_{\lambda}(\mathbf{k}),a_{\lambda^{\prime}}^{\dag}(\mathbf{k}^{\prime})
]=\delta_{\lambda\lambda^{\prime}}\delta^{(3)}(\mathbf{k}-\mathbf{k}^{\prime})$.
We define $h_{\lambda}(\tau,k)={u_k}(\tau)/{z_T}$ and $z_T=
a{ \sqrt{Q_T} }/2 $. Thus we get the equation of motion for $u_k$
as \be u_k^{\prime\prime}+\left(c_T^2
k^2-\frac{z_T^{\prime\prime}}{z_T} \right)u_k=0.\label{eom1} \ee

To phenomenologically investigate the effect on primordial GWs
spectrum induced by varying $c_T$, we assume that the variation of
$c_T$ can be described by a step-like function \ba
 c_T=\left\{ \begin{array}{ll}
c_{T1}\quad (\tau<\tau_0) \\ \label{ct12}
c_{T2}\quad (\tau>\tau_0)\\
\end{array}\right. ,
\ea where $\tau_0<0$ is the transition time.

We take the background evolution to be the slow-roll inflation. Of
course, it is also provided that the sudden change of $c_T$ won't
affect the background evolution.
Then, we have 
${z_T^{\prime\prime}}/{z_T}\equiv {a^{\prime\prime}}/{a}\approx
({2+3\epsilon})/{\tau^2}$, and the equation of motion (\ref{eom1})
becomes \be u_k^{\prime\prime}+\left(c_{T}^2
k^2-\frac{\nu^2-{1}/{4}}{\tau^2} \right)u_k=0,\label{eom2} \ee
where $\nu=\sqrt{\frac{9}{4}+3\epsilon}\approx
\frac{3}{2}+\epsilon $. The solution to Eq.(\ref{eom2}) is
familiar, we can write it as \ba u_{k1}&=&\sqrt{-c_{T1}k\tau}
\left[
C_{1,1}H^{(1)}_{\nu}(-c_{T1}k\tau)+C_{1,2}H^{(2)}_{\nu}(-c_{T1}k\tau)
\right],\quad \tau<\tau_0,\label{solution}
\nn\\
u_{k2}&=&\sqrt{-c_{T2}k\tau} \left[
C_{2,1}H^{(1)}_{\nu}(-c_{T2}k\tau)+C_{2,2}H^{(2)}_{\nu}(-c_{T2}k\tau)
\right],\quad \tau>\tau_0, \ea where $H^{(1)}_{\nu}$ and
$H^{(2)}_{\nu}$ are the first and second kind Hankel functions of
$\nu$-th order, respectively. These coefficients $C$ are
functions of the comoving wave number $k$, but constants with
respect to conformal time $\tau$. $C_{1,1}$ and $C_{1,2}$ are
determined by the initial condition.

Here, we set the initial condition as the standard
Bunch-Davies(BD) vacuum. Therefore, when $c_{T1}k\gg
\frac{a^{\prime\prime}}{a}$, which corresponds to perturbations
deep inside the horizon, \be
u_k\sim\frac{1}{\sqrt{2c_{T1}k}}e^{-ic_{T1}k\tau}.\label{bd} \ee
When $c_{T1}k\gg \frac{a^{\prime\prime}}{a}$, $u_{k1}$ in
Eq.(\ref{solution}) should approximate to Eq.(\ref{bd}). Allow for
the Hankel function $H^{(1)}_v(\xi)=\sqrt{\frac{2}{\pi
\xi}}e^{i(\xi-\frac{v\pi}{2}-\frac{\pi}{4})}$ and
$H^{(2)}_v(\xi)=\sqrt{\frac{2}{\pi
\xi}}e^{-i(\xi-\frac{v\pi}{2}-\frac{\pi}{4})}$ when
$|\xi|\rightarrow\infty$, we get \be
C_{1,1}=\frac{\sqrt{\pi}}{2\sqrt{c_{T1}k}},\,\,\,\,C_{1,2}=0. \ee

The coefficients $C_{2,1}$ and $C_{2,2}$ are determined by requiring $u_k$ and $u_k^{\prime}$ to be continuous at $\tau=\tau_0$, i.e. the matching condition. Then we obtain
\ba
C_{2,1}=\frac{  i \pi^{ \frac{3}{2} }\tau_0 k^{\frac{1}{2}}  }{16 \sqrt{ c_{T2} }   }
&\Big[&c_{T1}\left( H^{(1)}_{-1+\nu}(-c_{T1}k\tau_0)-H^{(2)}_{1+\nu}(-c_{T1}k\tau_0)  \right)H^{(2)}_{\nu}(-c_{T2}k\tau_0) \label{c1c2}
\nn\\
&+&c_{T2}\left( -H^{(2)}_{-1+\nu}(-c_{T2}k\tau_0)-H^{(2)}_{1+\nu}(-c_{T2}k\tau_0)  \right)H^{(1)}_{\nu}(-c_{T1}k\tau_0)
\,\Big],
\nn\\
C_{2,2}=\frac{  i \pi^{ \frac{3}{2} }\tau_0 k^{\frac{1}{2}}  }{16
\sqrt{ c_{T2} }   } &\Big[&c_{T1}\left(
-H^{(1)}_{-1+\nu}(-c_{T1}k\tau_0)+H^{(1)}_{1+\nu}(-c_{T1}k\tau_0)
\right)H^{(1)}_{\nu}(-c_{T2}k\tau_0)
\nn\\
&+&c_{T2}\left(
H^{(1)}_{-1+\nu}(-c_{T2}k\tau_0)-H^{(1)}_{1+\nu}(-c_{T2}k\tau_0)
\right)H^{(1)}_{\nu}(-c_{T1}k\tau_0) \,\Big]. \label{c21}\ea

The spectrum of primordial GWs is defined by
$P_T=({k^3}/{2\pi^2})\langle 0|\hat{h}_{ij}(\tau
,-\mathbf{k})\hat{h}_{ij}(\tau,\mathbf{k})  |0\rangle$
with $\tau\rightarrow 0$, which is only a function of comoving
wave number $k$.
After neglecting the slow-roll parameter, from Eq.(\ref{solution})
with $\nu=3/2$, we have \be
\left|u_{k2}\right|=\frac{\sqrt{2}}{-c_{T2}k\tau \sqrt{\pi}
}\left| C_{2,1}-C_{2,2} \right|. \ee Therefore, we obtain the
power spectrum of primordial GWs as \ba P_T & =&
\frac{k^3}{2 \pi^2}\sum_{\lambda=+,\times}|h_{\lambda}(\tau,k)|^2
=P^{inf}_T \frac{4 k}{Q_T \pi c_{T2}^2}\left|
C_{2,1}-C_{2,2}\right|^2,\label{spectrum} \ea where \be
P^{inf}_T=2H_{inf}^2/\pi^2 \label{Pgr}\ee is that of standard
slow-roll inflation without modified gravity, i.e. $Q_T=1$ and
$c_{T1}=c_{T_2}= 1$, and $H_{inf}$ is the Hubble parameter during
inflation, which sets the scale of inflation.

\begin{figure}[t!]
    \centering
\begin{minipage}[b]{0.48\linewidth}
    \centering
    \includegraphics[width=0.95\textwidth]{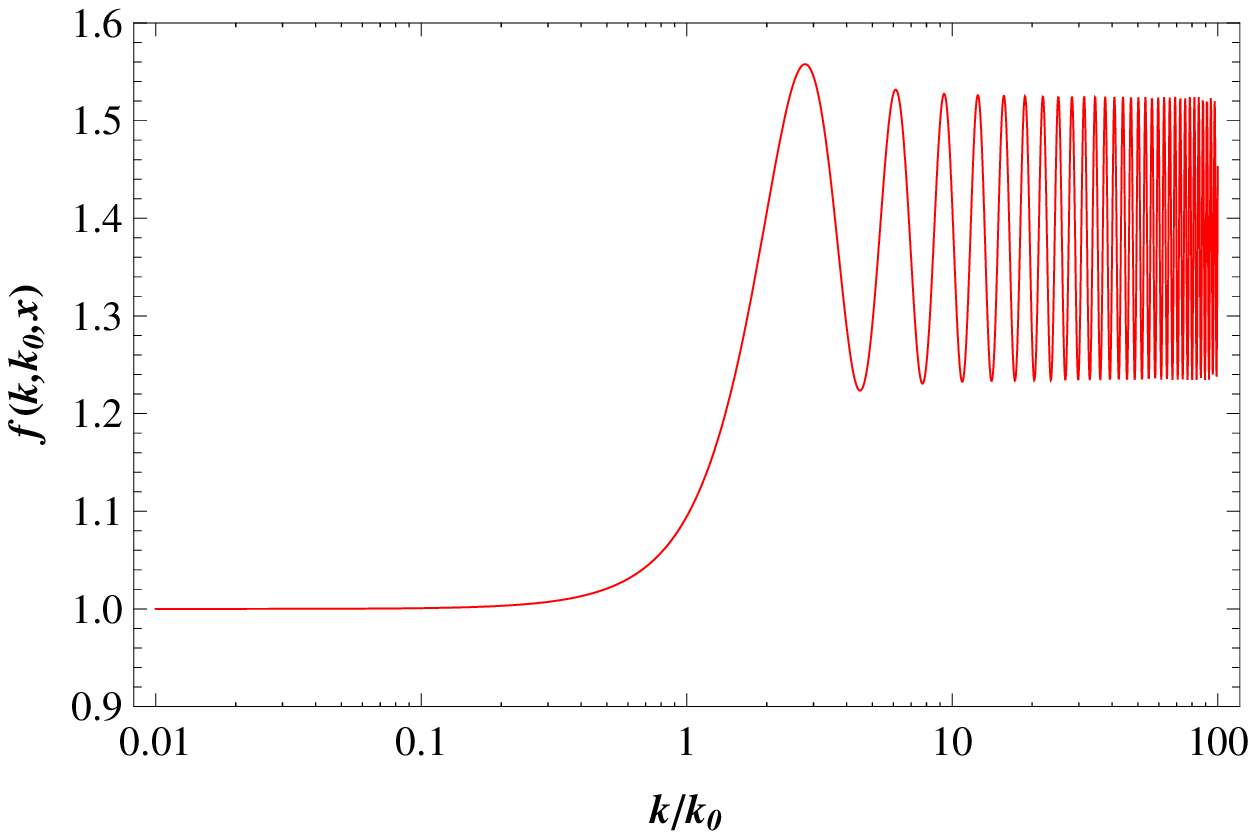}
    \end{minipage}
    \hspace{0.05cm}
\begin{minipage}[b]{0.48\linewidth}
    \centering
    \includegraphics[width=0.95\textwidth]{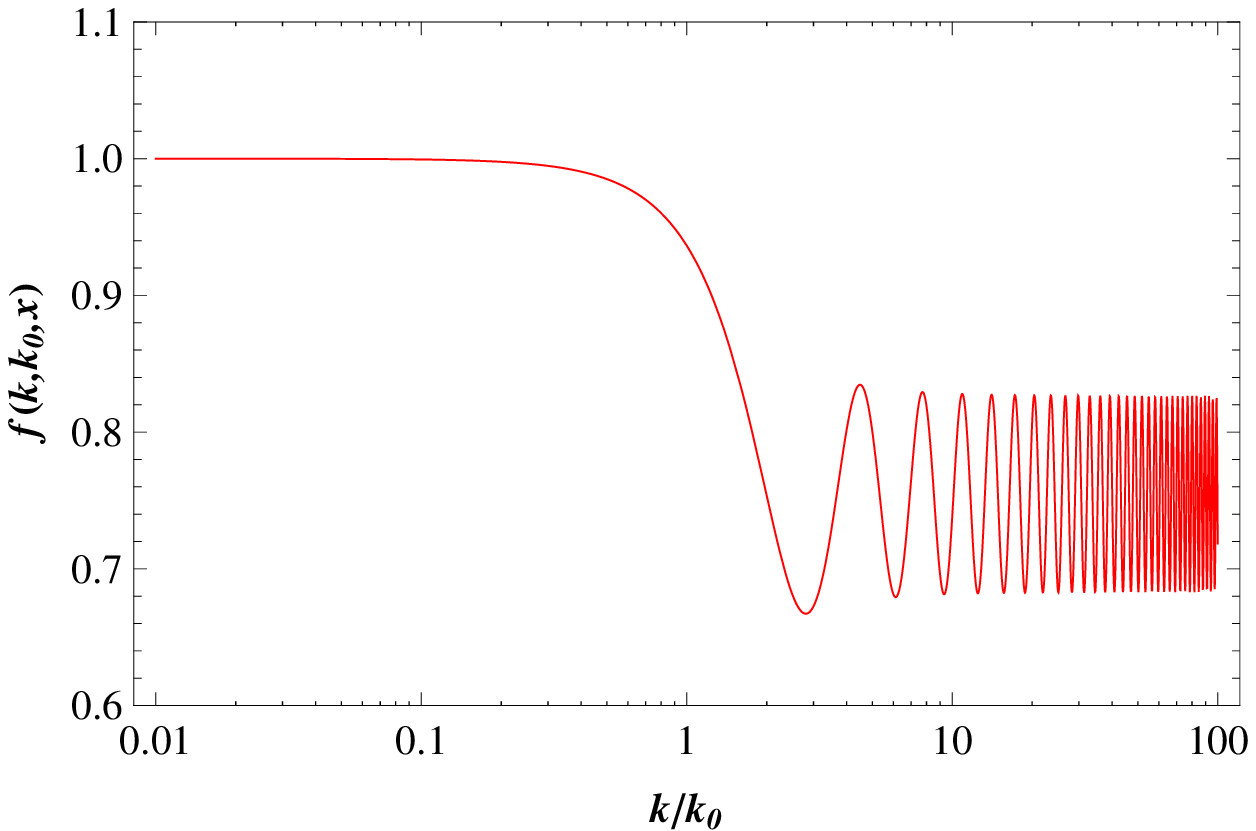}
    \end{minipage}
    \hspace{0.05cm}
\caption{The function $f(k,k_0,x)$, where $x=c_{T2}/c_{T1}$. We
have set $x=0.9$ on the left panel and $x=1.1$ on the right panel,
respectively.}
    \label{fig:v1}
\end{figure}

The effect of varying $c_T$ is encoded in $C_{2,1}$ and $C_{2,2}$.
We set $x= c_{T2}/c_{T1}$ and defined a new function \ba
f(k,k_0,x)=\frac{4 c_{T1} k }{\pi x^2}\left|
C_{2,1}-C_{2,2}\right|^2, \ea where $k_0=-1/(c_{T1} x \tau_0 )$.
Then, the GWs spectrum (\ref{spectrum}) may be rewritten as \be
P_T=P^{inf}_T \cdot\frac{f(k,k_0,x)}{ c_{T1}^3 Q_T },
\label{PTT}\ee where $f(k,k_0,x)$ is obtained as \ba f(k,k_0,x) =
\frac{1}{x^2} \sin^2\left(\frac{k}{k_0}\right)+\frac{1}{x^4}\left[
\cos(\frac{k}{k_0})-(1-x^2)\frac{k_0}{k}\sin(\frac{k}{k_0})
\right]^2.  \label{fk} \ea We plot $f(k,k_0,x)$ with respect to
$k/k_0$ in Fig.\ref{fig:v1}, in which we set $x=0.9$ on the left
panel and $x=1.1$ on the right panel, respectively. Here, the
transition time $\tau_0 =-1/(c_{T1} x k_0)$ sets a character scale
$1/k_0$. When $k\ll k_0$, i.e. the perturbation mode has longer
wavelength than $1/k_0$, we have $f(k,k_0,x)\approx 1$, and
$P_T=P^{inf}_T /{ (c_{T1}^3 Q_T) }$ is scale invariant, which is
the result of slow-roll inflation with almost constant $c_{T}$ and
$Q_T$ \cite{Kobayashi:2011nu}\cite{DeFelice:2011uc}. When $k\gg
k_0$, we have \be f(k,k_0,x)\approx \frac{1}{x^2}
\left[1+(\frac{1}{x^2}-1 )\cos^2(\frac{k}{k_0} ) \right],\ee thus
$f(k,k_0,x)$ oscillates between ${1}/{x^2}$ and ${1}/{x^4}$, and
$P_T$ oscillates correspondingly, just as we can see from
Fig.\ref{fig:v1}.

In the case of varying sound speed $c_S$ of primordial scalar
perturbations,
the sudden change of $c_S$ may lead to the oscillating modulation
to the primordial scalar spectrum, as well as the CMB TT-mode
spectrum, just as found in \cite{Nakashima:2010sa}.
In addition, the oscillation in the primordial scalar spectrum can
also be attributed to some other effects, such as inflaton
potential with a small oscillation
\cite{Wang:2002hf}\cite{Flauger:2009ab},
a sudden change in inflaton potential or its derivative, e.g.,
\cite{Starobinsky:1992ts}\cite{Adams:2001vc}\cite{Joy:2007na}\cite{Liu:2011cw}.
Thus the oscillation in the primordial scalar spectrum may be
implemented without modified gravity, as has been mentioned.


However, the oscillation in the primordial GWs spectrum can only
be attributed to the modified gravity. When the gravity is GR,
$P_T$ equals to $P^{inf}_T$, which is given in Eq.(\ref{Pgr}). The oscillation of $P^{inf}_T$
certainly requires $H_{inf}$ is oscillating, which is impossible,
unless the null energy condition is violated periodically. Though
the particle production may also modify the GWs spectrum
\cite{Cook:2011sp}\cite{Senatore:2011sp}\cite{Mukohyama:2014gba},
it only leads to a bump-like contribution, which is entirely
different from the behavior of oscillation.



\section{CMB B-mode polarization spectrum}

\begin{figure}[t!]
    \centering
\begin{minipage}[b]{0.48\linewidth}
    \centering
    \includegraphics[width=1.1\textwidth]{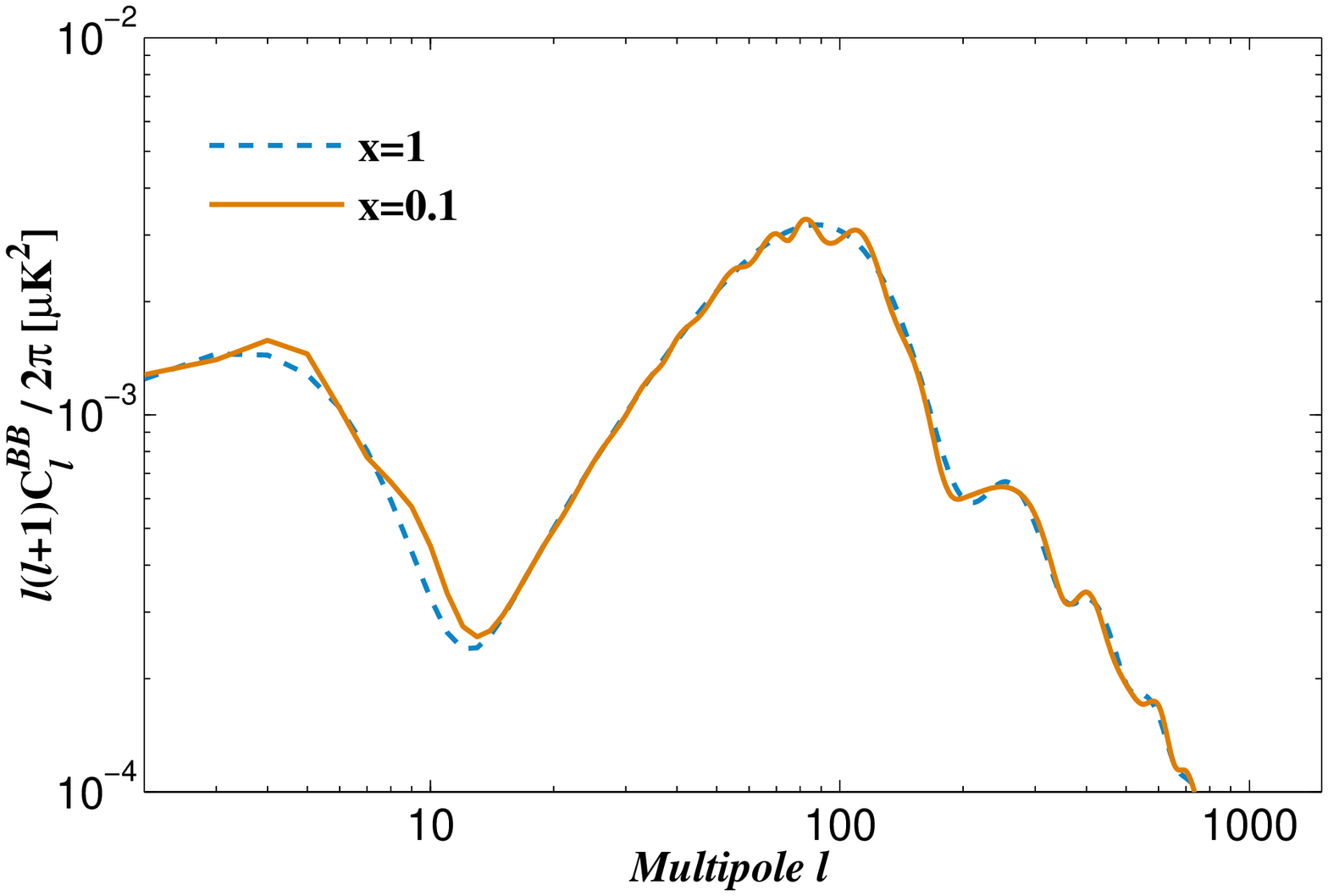}
    \end{minipage}
    \hspace{0.05cm}
\begin{minipage}[b]{0.48\linewidth}
    \centering
    \includegraphics[width=1.1\textwidth]{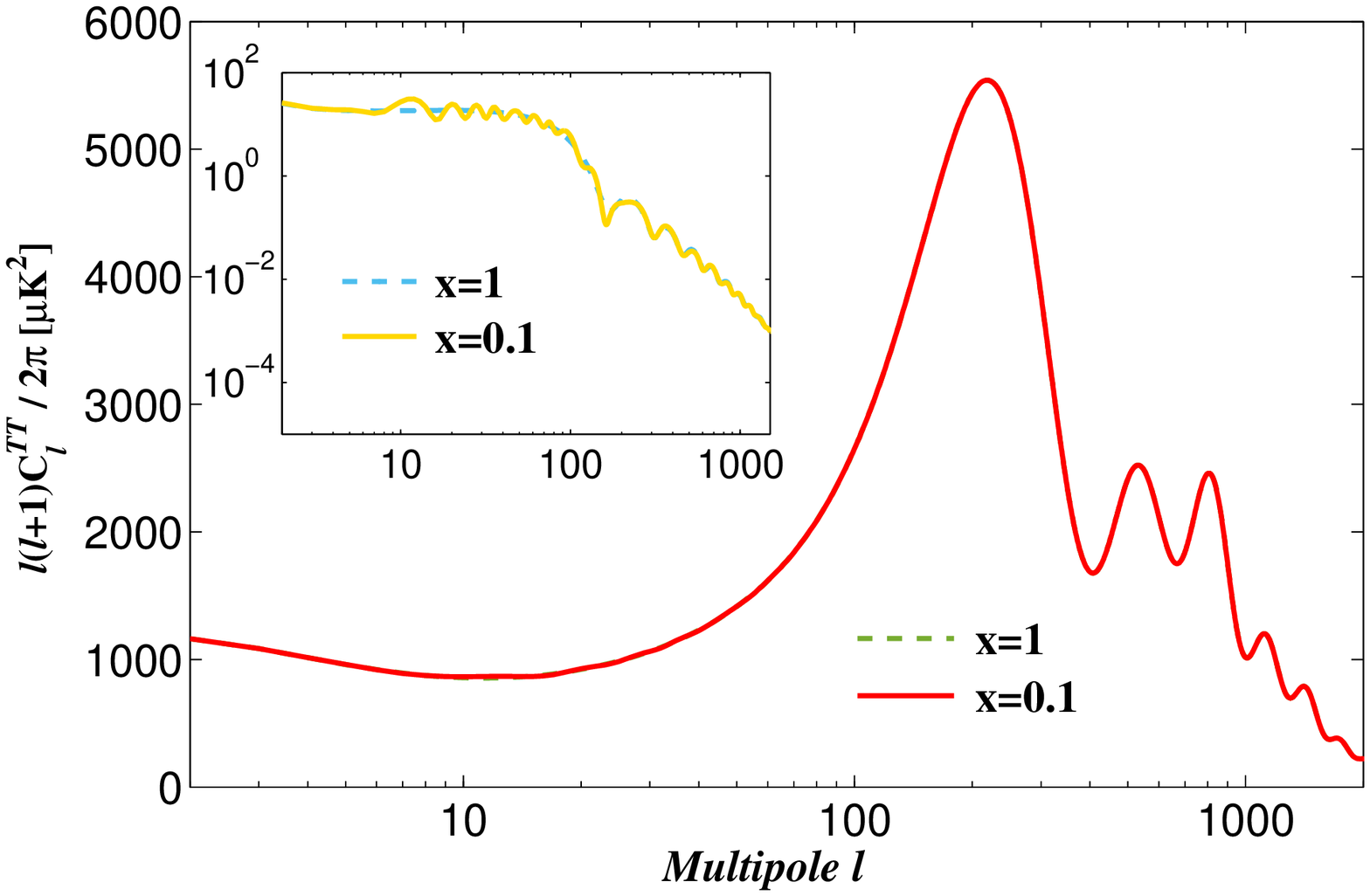}
    \end{minipage}
    \hspace{0.05cm}
\begin{minipage}[b]{0.48\linewidth}
    \centering
    \includegraphics[width=1.1\textwidth]{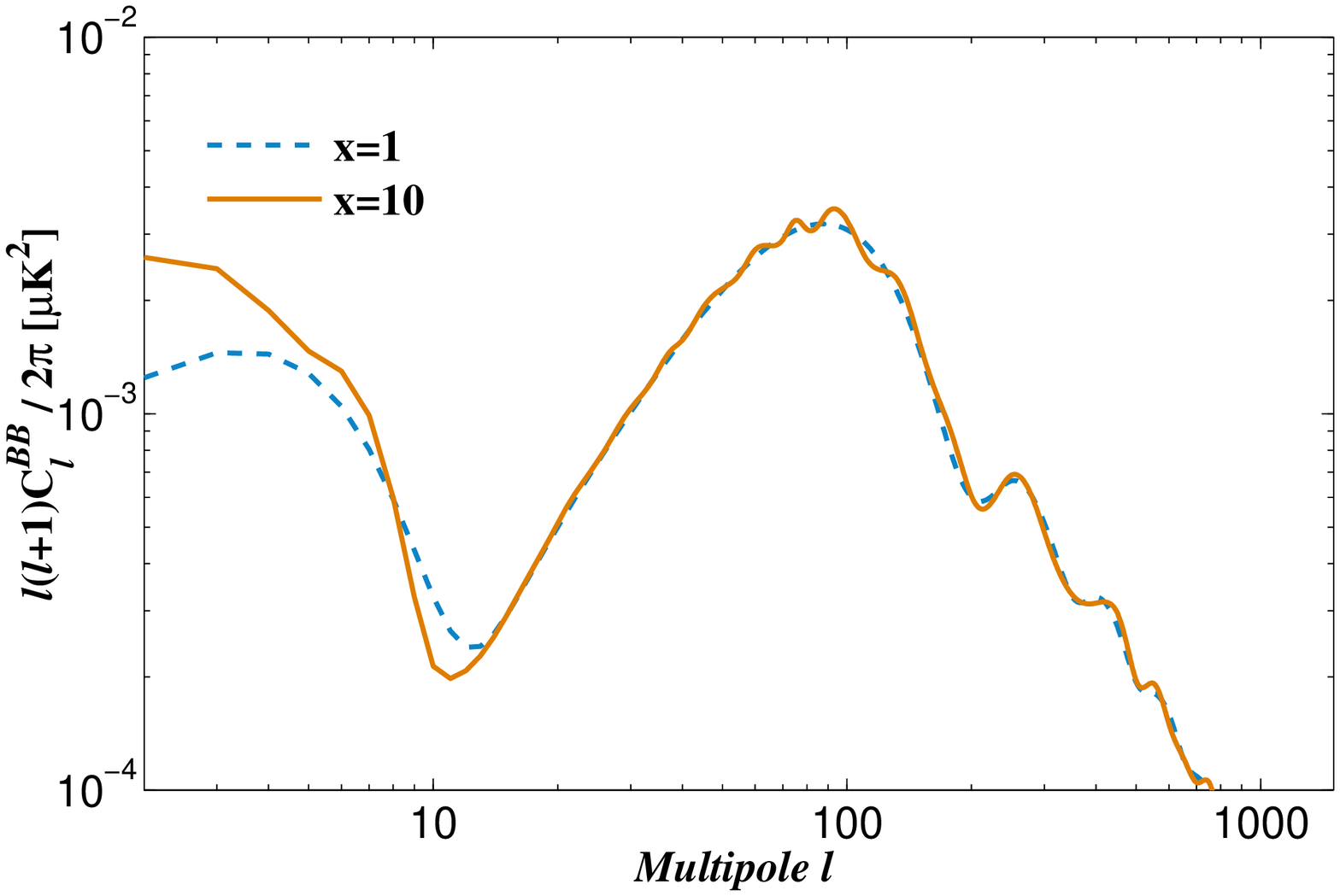}
    \end{minipage}
    \hspace{0.05cm}
\begin{minipage}[b]{0.48\linewidth}
    \centering
    \includegraphics[width=1.1\textwidth]{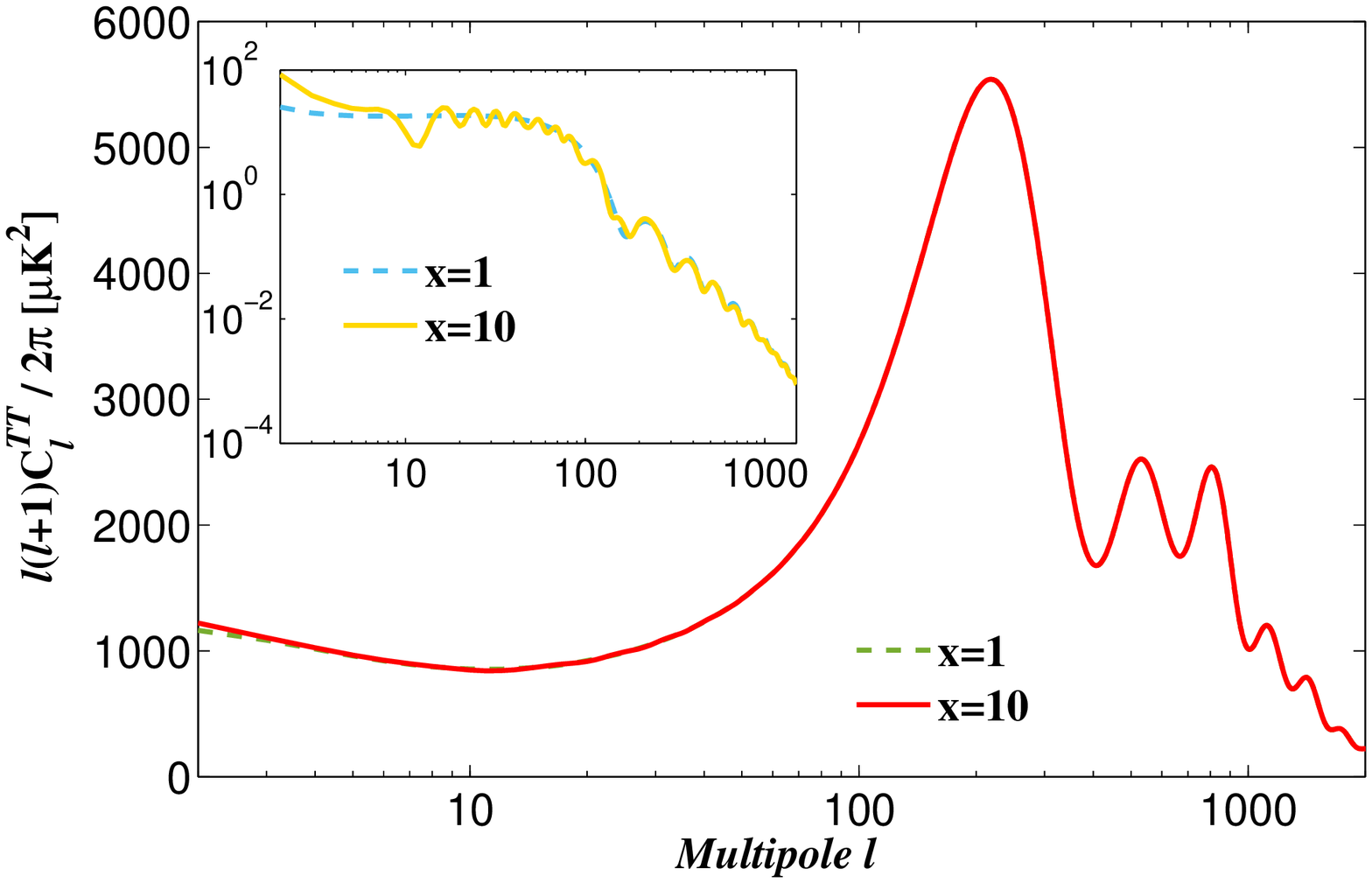}
    \end{minipage}
    \hspace{0.05cm}
\caption{Theoretical CMB BB and TT-mode power spectra for our
oscillating GWs spectrum (\ref{PTT})(brown line in the left panel and
red solid line in the right panel) and the power-law GWs spectrum for
reference(blue dashed line in the left panel and green dashed line in the
right panel). The insets of the right panels are the TT-mode spectra
for our oscillating GWs spectrum (the yellow solid lines) and the
power-law GWs spectrum (the blue dashed lines) for reference. We
set $r=0.05$ and $k_0=1/30000 \,\mathrm{Mpc}^{-1}$. } \label{fig:v2}
\end{figure}

The primordial GWs is imprint in CMB as the B-mode polarization.
Thus the oscillation in the primordial GWs spectrum will
inevitably affect the B-mode polarization spectrum.

To see such effects, we plot the CMB BB and TT-mode correlations
in Fig.\ref{fig:v2}, in which $ P^{inf}_T$ in (\ref{PTT}) is parameterized
as \be { P}^{inf}_T=rA_{{\cal R}}^{inf}\left({k\over
k_*}\right)^{n_{T}^{inf}}. \ee Here, we assume that the scalar
spectrum is hardly affected by the modified gravity, which will be
clarified in Appendix A. Thus the scalar perturbation spectrum is
set as ${ P}^{inf}_{\cal R}=A_{{\cal R}}^{inf}\left({k/
k_*}\right)^{n_{\cal R}^{inf}-1}$, in which $A_{{\cal R}}^{inf}$
is the amplitude of the scalar perturbations. In addition, we also
assume that after inflation the propagating speed $c_T$ is unity,
so that the spectrum of B-mode polarization is not affected by
relevant evolution at late time, or see
\cite{Amendola:2014wma},\cite{Raveri:2014eea}.



 \begin{figure}[t!]
    \centering
    \includegraphics[width=0.8\textwidth]{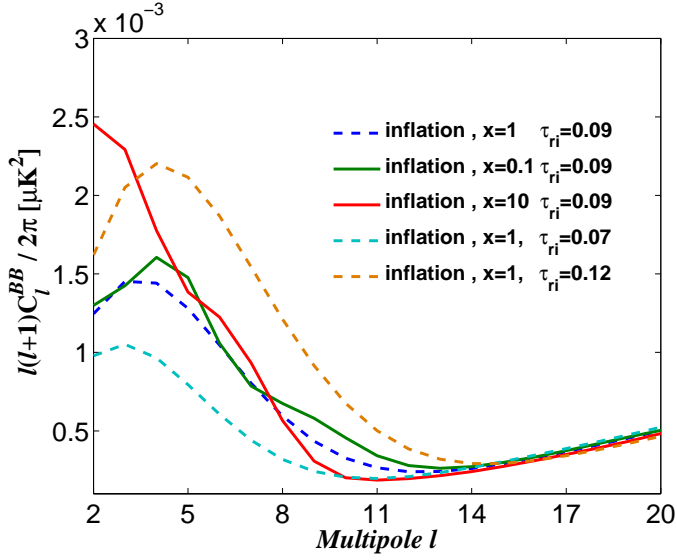}
    \hspace{0.05cm}
\caption{BB-mode spectra at low multipoles for our oscillating GWs
spectrum (\ref{PTT}) with different $x$(solid lines) and the
power-law GWs spectrum with different $\tau_{ri}$(dashed lines). }
    \label{fig:v3}
\end{figure}

In the left panels of Fig.\ref{fig:v2}, we see some obvious
enhancements or suppressions to the reionization bump in the BB-mode
spectrum, which depend on the oscillating effect on corresponding
scales. The height of the reionization bump can be estimated
roughly as \cite{Saito:2007kt}, \ba
C_{T,l\sim2}^{BB}&\approx&{1\over
100}\left(1-e^{-\tau_{ri}}\right)^2 C_{T,l\sim2}^{TT},
\label{tau}\ea where $C_{T,l}^{TT}$ stands for the TT-mode
spectrum from the primordial GWs without the reionization and
$\tau_{ri}$ is the optical depth to the beginning of reionization.
The periodic enhancements and suppressions of the reionization
bump are a reflection of the oscillations of primordial GWs
spectrum on large scales. In addition, we can also see some
obvious oscillations around the recombination peak at $l\sim 80$.

In the right panels of Fig.\ref{fig:v2}, we see that the TT-mode
spectrum is hardly affected by the oscillating primordial GWs
power spectrum, since the contribution of GWs to TT-mode spectrum
is negligible, compared with the scalar perturbations. The case of
EE-mode polarization spectrum is actually also similar. Thus the
main effect of varying speed of primordial GWs is on the B-mode
polarization, which makes the B-mode polarization spectrum show
its obvious enhancements or suppressions to the reionization bump
and oscillations around the recombination peak.

In Ref.\cite{Creminelli:2014wna}, the authors have pointed
out that it is possible to set to one the propagating speed of GWs
by a proper redefinition of the metric. They got the
gravitational waves spectrum in their Eq.(12) ``same" as the
standard one in the form. However, since they have made a
redefinition of the time coordinate and the scale factor in
Eq.(9), the variation of their $\tilde{H}$ with respect to $\tilde{t}$
after redefinition is not the same as the variation of $H$ with respect to
$t$. Therefore, the oscillating feature induced by a step-like
$c_T$ is encoded in $\tilde{H}$, the result in both frame should
be the same.

It might also be a concern whether such a B-mode polarization
spectrum can be imitated by adjusting other cosmological
parameters or not. In Eq.(\ref{tau}), the optical depth
$\tau_{ri}$ is relevant with the height of the reionization bump.
We show the BB-mode spectrum with different $\tau_{ri}$ in
Fig.\ref{fig:v3}. We see that the change of $\tau_{ri}$ can only
alter the overall amplitude of BB-mode spectrum at low multipoles,
but hardly create the oscillation. In addition, in
inflationary models with pre-inflation era, the reionization bump
could also be suppressed
(e.g.\cite{Piao:2003zm},\cite{Liu:2013kea},\cite{Liu:2013iha},\cite{Cai:2015nya})
or enhanced (e.g.\cite{Jain:2009pm}). However, similarly, these models also
only alter the overall amplitude of the BB-mode spectrum, without
oscillation, at low multipoles. These results indicate that
although the BB-mode spectrum may be modified by other ways, the
oscillating modulation leaded by varying the speed of primordial GWs
is difficult to be imitated. Thus the measure of B-mode
polarization spectrum provides an appropriate way for testing the
corresponding gravity physics in the primordial universe.


\section{Discussion}


In low-energy effective string theory and modified gravity
theories, the propagating speed $c_T$ of primordial GWs may
deviate from unity. We calculated the spectrum of primordial GWs,
assuming that $c_T$ has a step-like variation and the background
of slow-roll inflation is not affected by it. We found the
spectrum of primordial GWs acquires an oscillating modulation,
which makes the B-mode polarization spectrum show its obvious
enhancements or suppressions to the reionization bump and
oscillations around the recombination peak. The intensity of the
modulation is determined by $c_{T2}/c_{T1}$. The frequency of
the modulation is determined by $k_0=-1/(c_{T2} \tau_0 )$. Both
depend on the dynamics of theoretical models.

The oscillating behavior of the B-mode polarization can only be
attributed to the effect of modified gravity, since it can hardly
be imitated by adjusting other cosmological parameters. The
oscillating behavior is so rich in feature that it may be easily
discriminated from the variation of other parameters or other
features, thus the upcoming CMB experiments, such as CMBPol,
B-Pol, 
will be able to put a tight constraint on the
propagating speed of primordial GWs, and so the corresponding
theories, provided that the foreground contribution is under
control. In a certain sense, our paper again highlights the
significance of B-mode polarization measures in exploring the
fundamental physics of primordial universe.

Here, we only postulate a simple step-like variation of $c_T$,
which, however, might be far complicated in some modified gravity
models, as well as accompanied by the variation of $Q_T$.
The effect could be non-trivial in more general cases, which is
under study. But in a certain sense, a smooth change of $c_T$ will
induce oscillations too, see e.g.\cite{Achucarro:2014msa} for the
case of scalar perturbations, so we expect that the case of GWs is
similar. When we focus on the B-mode polarization
spectrum, the assumption we adopted is that 
after inflation the propagating speed $c_T$ is unity, which may
also be relaxed. Moreover, it may well be possible to
produce oscillatory features beyond the standard slow-roll
background. The varying $c_T$ and $Q_T$ will also affect the
non-Gaussianities of primordial perturbations. The relevant issues
are open.

\textbf{Acknowledgments}

This work is supported by NSFC, No. 11222546, and National Basic
Research Program of China, No. 2010CB832804. We acknowledge the use
of CAMB.

\section*{Appendix A:}

In this Appendix, we will argue how to realize the
change of $c_T$ and the changeless of $Q_T$ in (\ref{action2}) in
low-energy effective string theory and modified gravity theories.


In low-energy effective action of string theory, the simplest
extension of the lowest-order action is e.g.\cite{Cartier:2001is}
\be {\cal L}_{correction}\sim
-\frac{\alpha^{\prime}\lambda\xi(\varphi)}{2}\left( c_1
R_{GB}^2+c_2 G^{\mu\nu}\partial_{\mu}\varphi\partial_{\nu}\varphi
\right) \label{action3}\ee where
$G^{\mu\nu}=R^{\mu\nu}-g^{\mu\nu}R/2$, and
$R_{GB}^2=R_{\mu\nu\lambda\rho}R^{\mu\nu\lambda\rho}-4R_{\mu\nu}R^{\mu\nu}+R^2$
is the Gauss-Bonnet term, $\alpha^{\prime}$ is the inverse string
tension, $\lambda$ is a parameter allowing for different species
of string theories, $c_1$ and $c_2$ are coefficients. We have
neglected the terms with $\Box\varphi$ and $(\partial_\mu
\varphi\partial^{\mu}\varphi)^2$, since both do not contribute to
GWs.

The introducing of (\ref{action3}) will affect not only the GWs,
but also the adiabatic scalar perturbations, of course, it is
interesting to check its effect on the latter. However, for our
purpose, we will regard $\varphi$ as the spectator field, so that
the effects of (\ref{action3}) on the background and the scalar
perturbations are negligible.

The action for GWs is (\ref{action2}),
and \cite{Cartier:2001is} \ba
Q_T&=&1-\frac{\alpha^{\prime}\lambda}{2} \left( 8c_1 \dot{\xi}H_{inf}-c_2\xi \dot{\varphi}^2 \right),\nn\\
c_{T}^2&=&\frac{1}{Q_T}\left[
1-\frac{\alpha^{\prime}\lambda}{2}\left(c_2 \xi \dot{\varphi}^2 +
8c_1 \ddot{\xi} \right)  \right],  \ea see also
Ref.\cite{Feng:2013pba} for that with $c_1=0$. When $8c_1
\dot{\xi}H_{inf} \equiv c_2\xi \dot{\varphi}^2$ is imposed,
$Q_T=1$.
Then, we have
\be c_T^2=1-4c_1 \alpha^{\prime}\lambda(\dot{\xi}H_{inf}+\ddot{\xi}).
\ee
The step-like variation of
$c_T^2$ requires that $ \dot{\xi}H_{inf}+\ddot{\xi}$ has the
step at $\tau=\tau_0$.

\begin{figure}[t!]
    \centering
\begin{minipage}[b]{0.45\linewidth}
    \centering
    \includegraphics[width=0.95\textwidth]{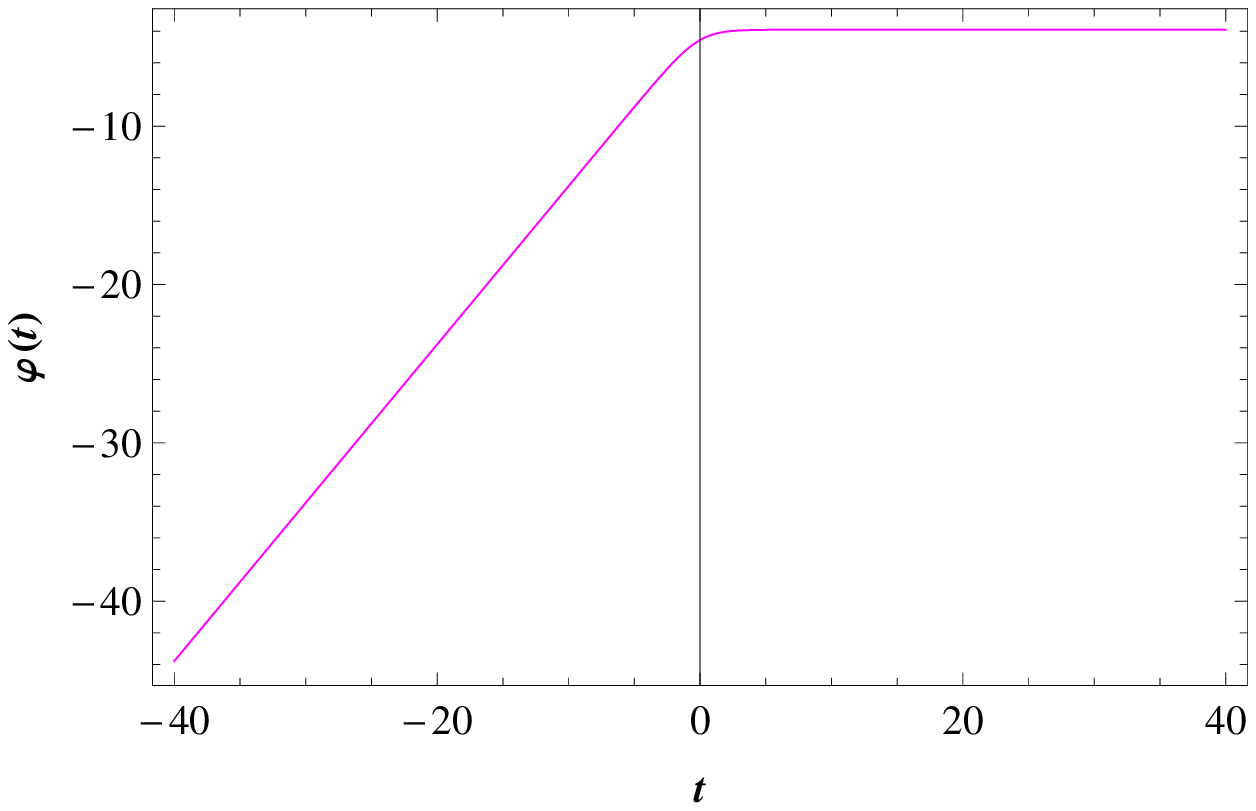}
    \end{minipage}
    \hspace{0.05cm}
\begin{minipage}[b]{0.45\linewidth}
    \centering
    \includegraphics[width=0.95\textwidth]{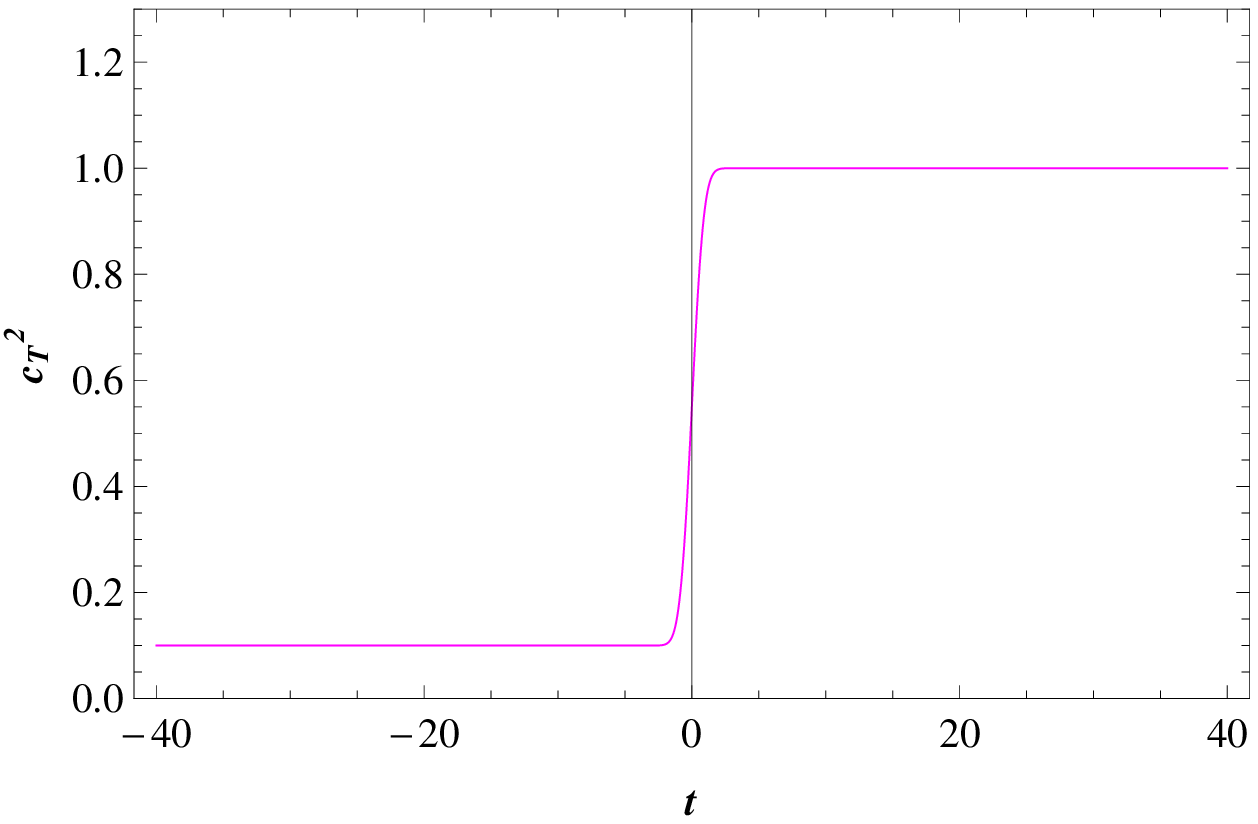}
    \end{minipage}
    \hspace{0.05cm}
\caption{The variation of $\varphi$ and $c_{T}^2$ with respect to cosmological time $t$ in the case of $x=c_{T2}/c_{T1}=10$, i.e., $A=-0.9$, where $t=0$ corresponds to $\tau_0$.  We have set $b_1=50$, $b_2=-50$.}
    \label{fig:v4}
\end{figure}

As an example, we will give a numerical result of the variation of $c_T^2$ and $\varphi$ with respect to cosmological time $t$ in the case of $x=c_{T2}/c_{T1}=10$. According to Refs.\cite{Met:1987}\cite{Cartier:2001is}, we adopt
\be \xi(\varphi)=-e^{-\varphi},\,\, c_1=-1,\,\, \alpha^{\prime}=1,\,\, \textrm{and}\, \lambda=-\frac{1}{8}\,(\textrm{for Heterotic\, string}).
\ee
We take the expression of $\varphi(t)$ as
\be \varphi(t)=t-\ln\left[b_1e^t+b_2A-Ae^t-\frac{Ae^{t-t^2}}{\sqrt{\pi} }+At e^t+Ae^{\frac{1}{4}}
\textrm{erf}(\frac{1}{2}-t)-Ae^t(t-1)\textrm{erf}(t)         \right],
\ee
where $A=-1+1/x$,  and $\textrm{erf} (t)=\frac{2}{\sqrt{\pi}}\int_0^te^{-t^2}\textrm{d}t$ is the Gauss error function, $b_1$ and $b_2$ are some constants. We plot $\varphi$ and $c_{T}^2$ with respect to cosmological time $t$ in Fig.\ref{fig:v4}.

Of course, the variation of $c_T^2$ could
be more complicated than a step-like evolution, which would be
harder to deal with. However, as it may, $Q_T>0$ and $c_T^2>0$
should be required to avoid ghost and gradient instabilities.

The action (\ref{action3}) is actually equivalent to a subclass of
the Horndeski theory, see Ref.\cite{Kobayashi:2011nu}, and so the
analysis is also similar for the Horndeski theory.


\begin{thebibliography}{99}

\bibitem{Gri:1974} L. P. Grishchuk, Sov.\ Phys.\ JETP {\bf 40},
409 (1975).


\bibitem{Sta:1979} A. A. Starobinsky, JETP Lett.\ {\bf 30}, 682
(1979).

\bibitem{Rub:1982} V. A. Rubakov, M. V. Sazhin, and A. V. Veryaskin, Phys.\ Lett.\ {\bf B115}, 189
(1982).

\bibitem{Ade:2014xna}
  P.~A.~R.~Ade {\it et al.}  [BICEP2 Collaboration],
  Phys.\ Rev.\ Lett.\  {\bf 112}, 241101 (2014)
  [arXiv:1403.3985 [astro-ph.CO]].

\bibitem{Adam:2014bub}
  R.~Adam {\it et al.}  [Planck Collaboration],
  arXiv:1409.5738 [astro-ph.CO].

\bibitem{Ade:2015tva}
  P.~A.~R.~Ade {\it et al.}  [BICEP2 and Planck Collaborations],
  [arXiv:1502.00612 [astro-ph.CO]].

\bibitem{Mortonson:2014bja}
  M.~J.~Mortonson and U.~Seljak,
  arXiv:1405.5857 [astro-ph.CO].

\bibitem{Ashoorioon:2012kh}
  A.~Ashoorioon, P.~S.~Bhupal Dev and A.~Mazumdar,
  Mod.\ Phys.\ Lett.\ A {\bf 29}, no. 30, 1450163 (2014)
  [arXiv:1211.4678 [hep-th]].








\bibitem{Grishchuk:2007uz}
  L.~P.~Grishchuk,
  In *Ciufolini, I. (ed.), Matzner, R.A. (ed.): General relativity and John Archibald Wheeler* 151-199
  [arXiv:0707.3319 [gr-qc]].




\bibitem{Lewis:2006fu}
  A.~Lewis and A.~Challinor,
  Phys.\ Rept.\  {\bf 429}, 1 (2006)
  [astro-ph/0601594].



\bibitem{Amendola:2014wma}
  L.~Amendola, G.~Ballesteros and V.~Pettorino,
  Phys.\ Rev.\ D {\bf 90}, 043009 (2014)
  [arXiv:1405.7004 [astro-ph.CO]].

\bibitem{Raveri:2014eea}
  M.~Raveri, C.~Baccigalupi, A.~Silvestri and S.~Y.~Zhou,
  arXiv:1405.7974 [astro-ph.CO].

\bibitem{Piao:2006ja}
  Y.~S.~Piao,
  Phys.\ Rev.\ D {\bf 75}, 063517 (2007)  [gr-qc/0609071];
  Phys.\ Rev.\ D {\bf 79}, 067301 (2009)  [arXiv:0807.3226 [gr-qc]].




\bibitem{Met:1987} R. R. Metsaev and A. A. Tseytlin, Nucl.\ Phys.\ {\bf
B293}, 385 (1987).

\bibitem{Antoniadis:1993jc}
  I.~Antoniadis, J.~Rizos and K.~Tamvakis,
  Nucl.\ Phys.\ B {\bf 415}, 497 (1994)  [hep-th/9305025].

\bibitem{Cartier:1999vk}
  C.~Cartier, E.~J.~Copeland and R.~Madden,
  JHEP {\bf 0001}, 035 (2000)
  [hep-th/9910169].

\bibitem{Cartier:2001is}
  C.~Cartier, J.~c.~Hwang and E.~J.~Copeland,
  Phys.\ Rev.\ D {\bf 64}, 103504 (2001)
  [astro-ph/0106197].

\bibitem{Piao:2003hh}
  Y.~S.~Piao, S.~Tsujikawa and X.~m.~Zhang,
  Class.\ Quant.\ Grav.\  {\bf 21}, 4455 (2004)  [hep-th/0312139].


\bibitem{Nojiri:2005jg}
  S.~Nojiri and S.~D.~Odintsov,
  Phys.\ Lett.\ B {\bf 631}, 1 (2005)  [hep-th/0508049];
  K.~Bamba, A.~N.~Makarenko, A.~N.~Myagky and S.~D.~Odintsov,
  Phys.\ Lett.\ B {\bf 732}, 349 (2014)  [arXiv:1403.3242 [hep-th]].


\bibitem{Horndeski:1974wa}
  G.~W.~Horndeski,
  Int.\ J.\ Theor.\ Phys.\  {\bf 10}, 363 (1974).

\bibitem{Amendola:1993uh}
  L.~Amendola,
  Phys.\ Lett.\ B {\bf 301}, 175 (1993)  [gr-qc/9302010].

\bibitem{Deffayet:2011gz}
  C.~Deffayet, X.~Gao, D.~A.~Steer and G.~Zahariade,
  Phys.\ Rev.\ D {\bf 84}, 064039 (2011)  [arXiv:1103.3260 [hep-th]].


\bibitem{Kobayashi:2011nu}
  T.~Kobayashi, M.~Yamaguchi and J.~Yokoyama,
  Prog.\ Theor.\ Phys.\  {\bf 126}, 511 (2011)
  [arXiv:1105.5723 [hep-th]].


\bibitem{Liu:2011ns}
  Z.~-G.~Liu, J.~Zhang and Y.~-S.~Piao,
  Phys.\ Rev.\ D {\bf 84}, 063508 (2011)  [arXiv:1105.5713
  ].

\bibitem{Gleyzes:2014dya}
  J.~Gleyzes, D.~Langlois, F.~Piazza and F.~Vernizzi,
  arXiv:1404.6495 [hep-th].

\bibitem{Gao:2014soa}
  X.~Gao,
  Phys.\ Rev.\ D {\bf 90}, 081501 (2014)  [arXiv:1406.0822
  [gr-qc]]; 
  X.~Gao,
  Phys.\ Rev.\ D {\bf 90}, 104033 (2014)  [arXiv:1409.6708 [gr-qc]].


\bibitem{Ballesteros:2014sxa}
  G.~Ballesteros,
  arXiv:1410.2793 [hep-th].






\bibitem{DeFelice:2014bma}
  A.~De Felice and S.~Tsujikawa,
  arXiv:1411.0736 [hep-th].

\bibitem{Nakashima:2010sa}
  M.~Nakashima, R.~Saito, Y.~i.~Takamizu and J.~Yokoyama,
  Prog.\ Theor.\ Phys.\  {\bf 125}, 1035 (2011)
  [arXiv:1009.4394 [astro-ph.CO]].

\bibitem{Firouzjahi:2014fda}
  H.~Firouzjahi and M.~H.~Namjoo,
  Phys.\ Rev.\ D {\bf 90}, 063525 (2014)
  [arXiv:1404.2589 [astro-ph.CO]].



\bibitem{Dubovsky:2010pe}
  S. Dubovsky, R. Flauger, A. Starobinsky and I. Tkachev,
  Phys.\ Rev.\ D {\bf 81}, 023523 (2010)
  [arXiv:0907.1658 [astro-ph.CO]].

\bibitem{Gumrukcuoglu:2012wt}
  A.~E.~Gumrukcuoglu, S.~Kuroyanagi, C.~Lin, S.~Mukohyama and N.~Tanahashi,
  Class.\ Quant.\ Grav.\  {\bf 29}, 235026 (2012)
  [arXiv:1208.5975 [hep-th]].



\bibitem{Huang:2012pe}
  Q.~G.~Huang, Y.~S.~Piao and S.~Y.~Zhou,
  Phys.\ Rev.\ D {\bf 86}, 124014 (2012)
  [arXiv:1206.5678 [hep-th]].

\bibitem{DeFelice:2011uc}
  A.~De Felice and S.~Tsujikawa,
  Phys.\ Rev.\ D {\bf 84}, 083504 (2011)  [arXiv:1107.3917 [gr-qc]].




















\bibitem{Wang:2002hf}
  X.~Wang, B.~Feng, M.~Li, X.~L.~Chen and X.~Zhang,
  Int.\ J.\ Mod.\ Phys.\ D {\bf 14}, 1347 (2005)  [astro-ph/0209242].


\bibitem{Flauger:2009ab}
  R.~Flauger, L.~McAllister, E.~Pajer, A.~Westphal and G.~Xu,
  JCAP {\bf 1006}, 009 (2010)  [arXiv:0907.2916 [hep-th]].


\bibitem{Starobinsky:1992ts}
  A.~A.~Starobinsky,
  JETP Lett.\  {\bf 55}, 489 (1992)
  [Pisma Zh.\ Eksp.\ Teor.\ Fiz.\  {\bf 55}, 477 (1992)].

\bibitem{Adams:2001vc}
  J.~A.~Adams, B.~Cresswell and R.~Easther,
  Phys.\ Rev.\ D {\bf 64}, 123514 (2001)
  [astro-ph/0102236].



\bibitem{Joy:2007na}
  M.~Joy, V.~Sahni and A.~A.~Starobinsky,
  Phys.\ Rev.\ D {\bf 77}, 023514 (2008)
  [arXiv:0711.1585 [astro-ph]];
  M.~Joy, A.~Shafieloo, V.~Sahni and A.~A.~Starobinsky,
  JCAP {\bf 0906}, 028 (2009)
  [arXiv:0807.3334 [astro-ph]].

\bibitem{Liu:2011cw}
  J.~Liu and Y.~S.~Piao,
  Phys.\ Lett.\ B {\bf 705}, 1 (2011)  [arXiv:1106.5608 [hep-th]].

\bibitem{Cook:2011sp}
  J. L. Cook, and L. Sorbo,
  Phys.\ Rev.\ D {\bf 85}, 023534 (2012)
  [arXiv:1109.0022 [astro-ph.CO]].

\bibitem{Senatore:2011sp}
  L.~Senatore, E.~Silverstein and M.~Zaldarriaga,
  JCAP {\bf 1408}, 016 (2014)  [arXiv:1109.0542 [hep-th]].

\bibitem{Mukohyama:2014gba}
  S.~Mukohyama, R.~Namba, M.~Peloso and G.~Shiu,
  JCAP {\bf 1408}, 036 (2014)  [arXiv:1405.0346 [astro-ph.CO]].


\bibitem{Saito:2007kt}
  S.~Saito, K.~Ichiki and A.~Taruya,
  JCAP {\bf 0709}, 002 (2007)
  [arXiv:0705.3701 [astro-ph]].

\bibitem{Creminelli:2014wna}
  P.~Creminelli, J.~Gleyzes, J.~Nore?a and F.~Vernizzi,
  Phys.\ Rev.\ Lett.\  {\bf 113}, no. 23, 231301 (2014)
  [arXiv:1407.8439 [astro-ph.CO]].




\bibitem{Piao:2003zm}
  Y.~-S.~Piao, B.~Feng and X.~-m.~Zhang,
  Phys.\ Rev.\ D {\bf 69}, 103520 (2004)  [hep-th/0310206];

\bibitem{Liu:2013kea}
  Z.~G.~Liu, Z.~K.~Guo and Y.~S.~Piao,
  Phys.\ Rev.\ D {\bf 88}, 063539 (2013)
  [arXiv:1304.6527]; 
  Y.~T.~Wang and Y.~S.~Piao,
  Phys.\ Lett.\ B {\bf 741}, 55 (2015) [arXiv:1409.7153 [gr-qc]].

\bibitem{Liu:2013iha}
  Z.~G.~Liu, Z.~K.~Guo and Y.~S.~Piao,
  Eur.\ Phys.\ J.\ C {\bf 74}, 3006 (2014)  [arXiv:1311.1599 [astro-ph.CO]].

\bibitem{Cai:2015nya}
  Y.~Cai, Y.~T.~Wang and Y.~S.~Piao,
  arXiv:1501.01730 [astro-ph.CO].

\bibitem{Jain:2009pm}
  R. K. Jain, P. Chingangbam, L. Sriramkumar and T. Souradeep,
  Phys. Rev. D 82, 023509 (2010) [arXiv:0904.2518 [astro-ph.CO]].

\bibitem{Achucarro:2014msa}
  A.~Achucarro, V.~Atal, B.~Hu, P.~Ortiz and J.~Torrado,
  Phys.\ Rev.\ D {\bf 90}, no. 2, 023511 (2014)
  [arXiv:1404.7522 [astro-ph.CO]].

\bibitem{Feng:2013pba}
  K.~Feng, T.~Qiu and Y.~S.~Piao,
  Phys.\ Lett.\ B {\bf 729}, 99 (2014)
  [arXiv:1307.7864 [hep-th]].































\end{thebibliography}
\end{document}